\documentclass[a4paper]{jpconf}
\usepackage{graphicx}
\begin{document}
\title{FPGA based data acquisition system for\newline \mbox{COMPASS} experiment}

\author{ M Bodlak$^1$, V Frolov$^3$$^4$,V Jary$^1$, S Huber$^2$, I Konorov$^2$, D Levit$^2$, J Novy$^1$, S Paul$^2$, R Salac$^1$ and M Virius$^1$}

\address{$^1$Faculty of Nuclear Sciences and Physical Engineering, Czech Technical University, Brehova 7, 115 19 Prague 1, Czech Repulic}
\address{$^2$Physik-Department E18, Technische Universitat Munchen, James-Franck-Straße, 85748 Garching, Germany}
\address{$^3$Joint Institute for Nuclear Research,
Joliot-Curie 6, 141980 Dubna, Moscow region, Russia}
\address{$^4$European Organization for Nuclear Research - CERN,
CH-1211, Geneve 23, Switzerland}

\ead{josef.novy@cern.ch}

\begin{abstract}
This paper discusses the present data acquisition system (DAQ) of the COMPASS experiment at CERN and presents development of a new DAQ. The new DAQ must preserve present data format and be able to communicate with FPGA cards.  Parts of the new DAQ are based on state machines and they are implemented in C++ with usage of the QT framework, the DIM library, and the IPBus technology. Prototype of the system is prepared and communication through DIM between parts was tested. An implementation of the IPBus technology was prepared and tested. The new DAQ proved to be able to fulfill requirements.
\end{abstract}

\section{Introduction}

This paper presents design, implementation, and preliminary performance results of a new data acquisition system (DAQ), based on the Field Programmable Gate Array (FPGA) technology, for the COMPASS experiment. COMPASS~\cite{COMPASS} is a fixed target experiment at CERN with a usual data rate of approximately 2000\,MB/s during 10~second on-spill with 45~second off-spill. Its present DAQ system was built during years 1999-2001 and now is ready for modernization with the FPGA technology. Since the FPGA technology started to be cost-effective and developed enough in recent years, the usage for event building purposes is now solvable at reasonable cost, and thus the new FPGA based DAQ can be designed, analysed, implemented, and deployed.

The Data Acquisition and Test Environment(DATE)~\cite{DATE} software package and usage of FPGA-based cards has been widely studied. Thanks to these studies, development of the new DAQ did not start from zero, but more likely took advantage of work already done in the field of DAQ software development. The development aim is to bring DAQ design closer to state of the art software and hardware designs, and thus improve performance and reliability. Improved reliability of the system should be achieved thanks to the movement of event building to hardware and more extensive usage of the OOP.

In the last few years a growing number of tools, which use Distributed Information Management System (DIM), were made for the COMPASS experiment, and as consequence usage of DIM package for communication was recommended. State machines used in the present DAQ were analyzed and reworked for use as a basis for main processes of the new DAQ.

\section{Used technologies}

In order to develop a new DAQ, we started by studying the present one, which is based on the DATE software package originally developed for the ALICE experiment at CERN. It is written in the C programming language and it uses the DIM library for communication. The DIM is a multi-platform library that serves for an asynchronous 1~to~many communication through the Ethernet. Extensive state machines implemented in State Management Interface (SMI++) are part of this system. Both DIM and SMI++ were originally developed for the DEPHI experiment at CERN. State machines in the present DAQ were taken as a basis for the new DAQ state machines but were significantly reworked to take into account the new DAQ architecture and implemented in the C++ programming language.

The Qt framework is a cross-platform application framework which is used both for the development of graphical user interfaces (GUI) and for non-GUI programs. Development of the Qt toolkit by the Trolltech company started in 1999. Now it is developed by Digia. The Qt framework is distributed under GNU LGPL, GNU GPL, and Commercial Developer License copyright licenses. The Qt framework was chosen originally as a tool to easily make GUI, but its role in design changed soon after start of the development. Now it is used in all parts of the new DAQ designs.  It is very useful thanks to its extensive support for multi-threading, databases (SQL), XML processing, file access, work with multimedia, and possibility to easily deploy developed software on a wide range of platforms such as: Windows, Symbian, Mac OS X, X11, Embedded Linux etc. The standard Qt SDK includes also development environment called Qt Creator. The Qt Creator consists of code editor, distributed revision control, and performance testing tools like the Valgrind and the Cachegrind. Subversion was used for version control.

The FPGA technology is key part of the new DAQ design. FPGA chips are special integrated circuits whose behavior can be changed in the field by uploading a new firmware. Usability of this technology has grown in the past few years, because chip density increased and the cost per programmable block dropped. This technology provides very good flexibility, which is very important in usage for physical experiments.

The IPBus package is used for communication with FPGAs. It was developed for the level one trigger update of the CMS experiment. This package consists of firmware part and software part. The firmware part mediates access to registers and memory of a FPGA card through Ethernet when properly loaded. The software part is implemented in C++ and contains all classes needed for a connection to the interface of the firmware~part.

\section{COMPASS DAQ architecture}

\subsection{Old DAQ architecture}

\begin{figure}[p]
    \begin{center}
        \includegraphics[width=16.0cm]{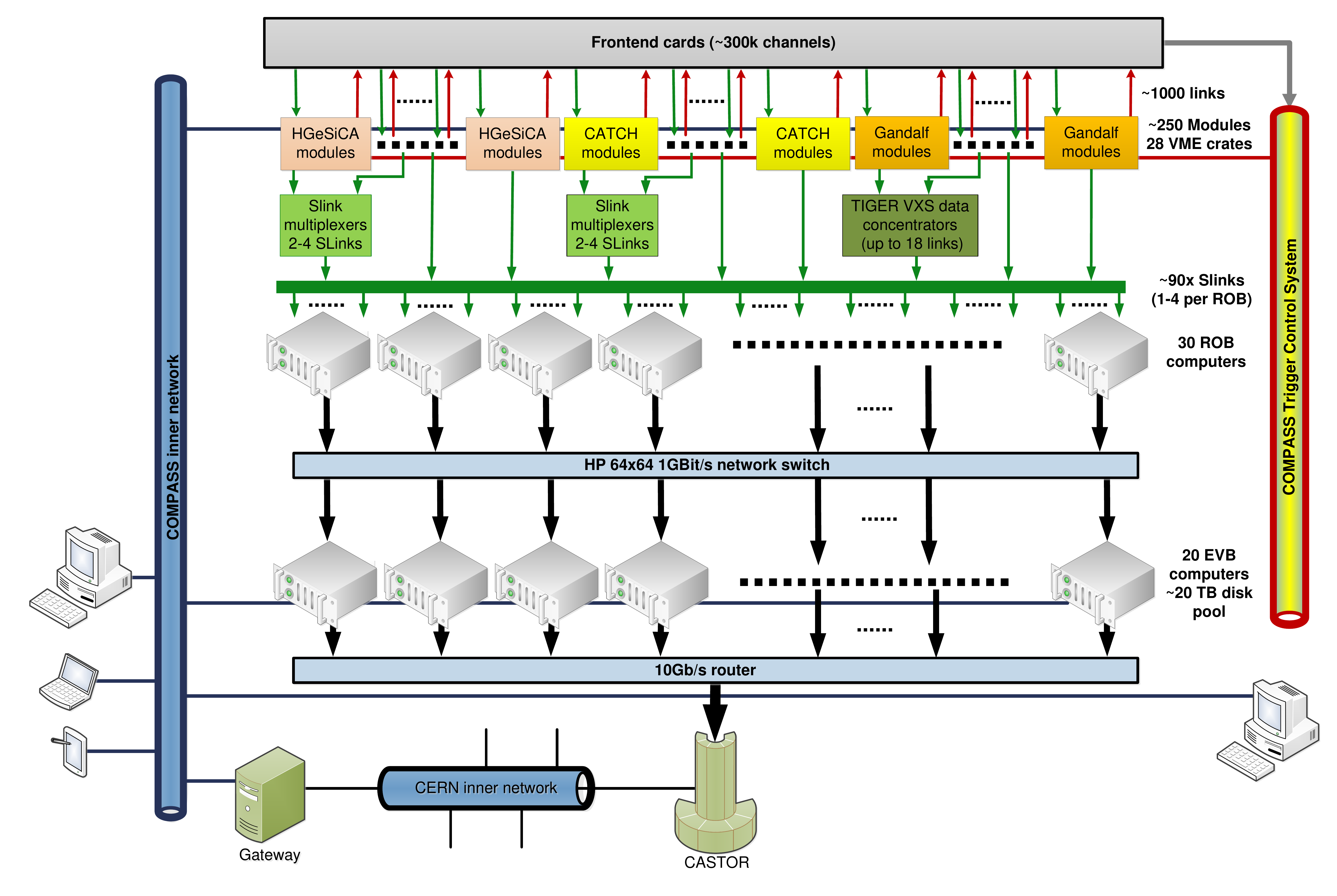}
        \caption{The old DAQ architecture} 
        \label{obr-old} 
        \includegraphics[width=16.0cm]{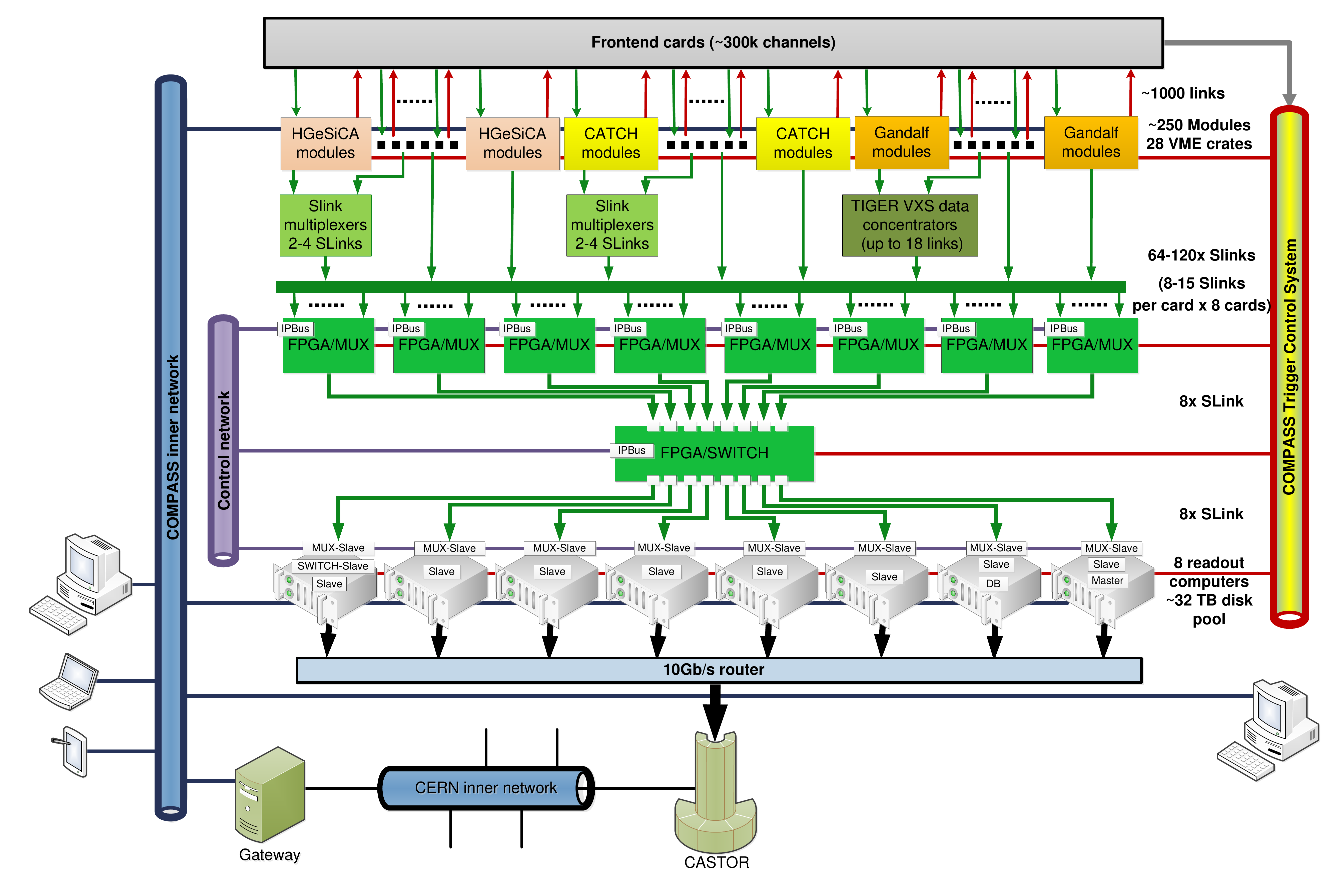}
        \caption{The new DAQ architecture} 
        \label{obr-arch} 
    \end{center}
\end{figure}

For good understanding of the changes in the new DAQ, it is necessary to start with a description of the present DAQ. Its basic structure is shown in Figure~\ref{obr-old}. It consists of three main parts: the data taking part, the event building part, and the control part. The data taking part incorporates detectors, frontend electronics, and data preprocessing. The event building part incorporates thirty readout buffer computers (ROBs) and twenty computers for the event building (EB) all interconnected through the Ethernet switch. The control part is then built mainly from ordinary PCs with the Scientific Linux CERN, the DATE package, and several monitoring utilities (COOOL, MurphyTV, etc.).

\subsection{New DAQ architecture}
 
\begin{figure}[h!t]
    \begin{center}
        \includegraphics[width=14.0cm]{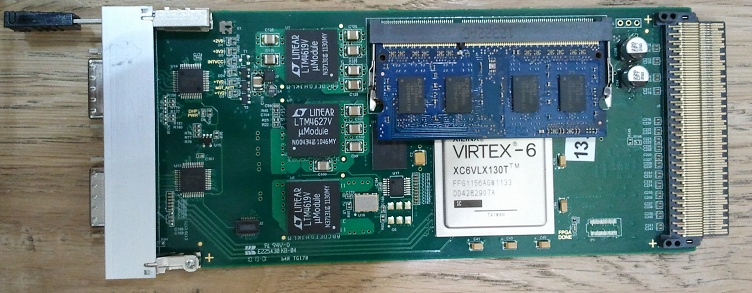}
        \caption{New Switch/Mux FPGA card} 
        \label{obr-FPGA} 
    \end{center}
\end{figure}

An idea of the new DAQ is based on the reworked event building part as shown in Figure~\ref{obr-arch}. The data taking part is the same as in the present DAQ. The event building part, on the other hand, is changed to incorporate new FPGA cards shown in Figure~\ref{obr-FPGA}. These cards will be used with two different roles in two layers. The first layer of FPGAs has the role of multiplexers. The second layer consists of only one card, which fulfills the event building role. It prepares full events in a special format. Events are then sent to readout computers, where they are readout, transformed to the DATE format, and stored temporarily on local disks, before transfer to the CERN Advanced STORage manager (CASTOR).

The reworked design of the event building part allowed using a smaller control system. Additionally, the system can also utilize more computing power for monitoring thanks to movement of the event building to FPGAs. The last but not the least important change is usage of a new design of state machines, which introducing second run state with tolerance to errors.

\subsection{Processes in the new DAQ}

\begin{figure}[h!t]
    \begin{center}
        \includegraphics[width=16.0cm,]{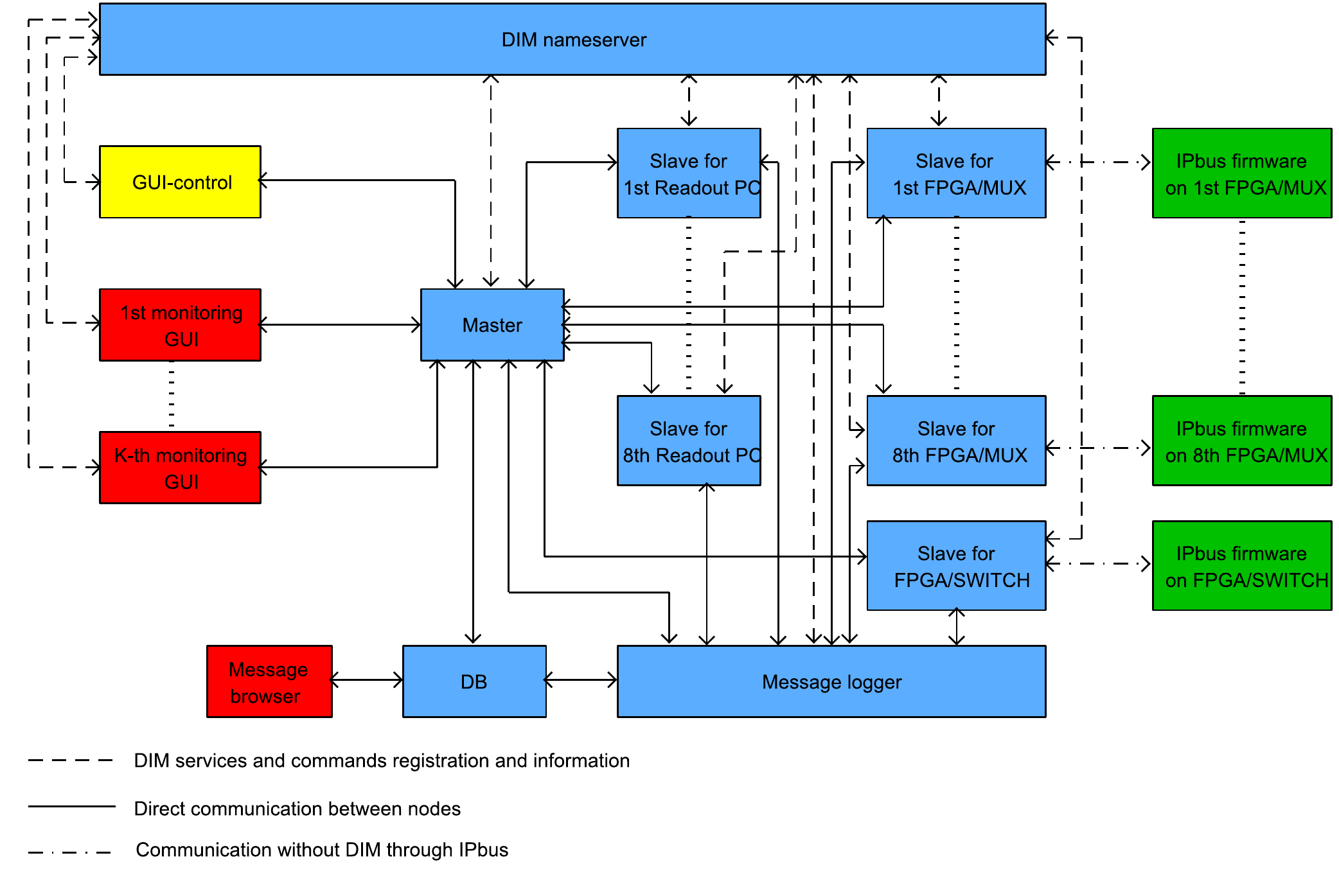}
        \caption{Communication diagram} 
        \label{obr-com} 
    \end{center}
\end{figure}

The software part of the new DAQ consists of Master, Slave-control, Slave-readout, Runcontrol GUI, MessageLoger, and MessageBrowser. The Master process acts as the middleman between a user interface and Slave processes. All communication is done through the DIM. The Master process keeps track of states of all Slave processes. The expected message rate per slave is 10-100\,Hz. Furthermore, the Master process has incorporated the main part of error handling and a connection to a configuration database. If the Slave process needs any information from the database, it must send a request to the Master process. All software parts and connections between them are shown in Figure~\ref{obr-com}.

The Slave-control process fulfills a role of the supervisor for FPGA cards. It uses the IPBus technology to gain access to registers and memory of the FPGA card. Its task is to configure the FPGA, keep track of FPGA’s resources, and inform the Master process about the state of the connected FPGAs. It can run on any computer connected to same network as the FPGA card, which it should control, and to the network with a computer on which the Master process runs. It is expected that it will run on one of the readout computers and that for interconnection of readout computers and FPGA cards a dedicated network will be used.

The Slave-readout process is the largest and the most resource demanding process in the~new~DAQ. It is a multithread process that serves for control of readout activities. The main threads are: informer, readout, FIFO, processor.

The informer thread uses a set of DIM services and commands for communication with the Master process. It sends information about a state of the whole Slave-readout process and processes commands.

The readout thread is responsible for readout of connected devices and preparation of readout data. All data sources are derived from a common abstract class; thus it is easy to add a new type of device to the system. This construction significantly increases flexibility of the system. An SLink Spillbuffer card is used as the only data source in the present prototype. The Spillbuffer card is an FPGA based PCI-Express card that serves for buffering of a data stream from data producers before loading them to a computer memory. The readout thread is also responsible for arming and disarming data sources. Checks of data consistency are also made during readout.

The FIFO thread is responsible for storing events in the DATE format, before they are processed. Transfers of events from the readout thread to the FIFO thread and from the FIFO thread to the processor thread are made through signal-slot connections, and they are done by blocks of about 1000 events. These block transfers allow reduction of the computer resource cost of data exchanges between different threads.

The main purpose processor thread’s is to store date on local disks, additionally it also distributes events based on the set filter to other outputs. Four basic outputs are defined in the~new~DAQ prototype: FileOutput, ScreenOutput, MonitoringOutput, and TerminatorOutput. All these outputs have their own threads, but only the FileOutput can pause processing when it cannot process data fast enough. The TerminatorOutput’s role is taking care of disposing events after all other outputs are done with them. The ScreenOutput and the MonitoringOutput serve for monitoring purposes.

The GUI process as the only one visible to standard users is the most complicated one from the design point of view. Its structure is designed and developed with great emphasis on ergonomics and flexibility. It provides different sets of information for a normal user and an expert user. It is prepared as fully customizable.  It can run in two modes: runcontrol and monitoring. Only one runcontrol GUI is allowed in the system for a better control of the system behavior. The number of running monitoring GUIs is not limited.

The MessageBrowser and the MessageLogger support programs~\cite{BDipl} are the last part of the~new~DAQ software design. The MessageLogger’s task is to receive all messages from all parts of the system and store them in the database. MessageBrowser is a visualization tool for browsing through these messages. It has extensive filtering capabilities that allow to find quickly the desired messages.

\section{Tests of prototype}

\begin{figure}[h!t]
    \begin{center}
        \includegraphics[width=13.0cm,]{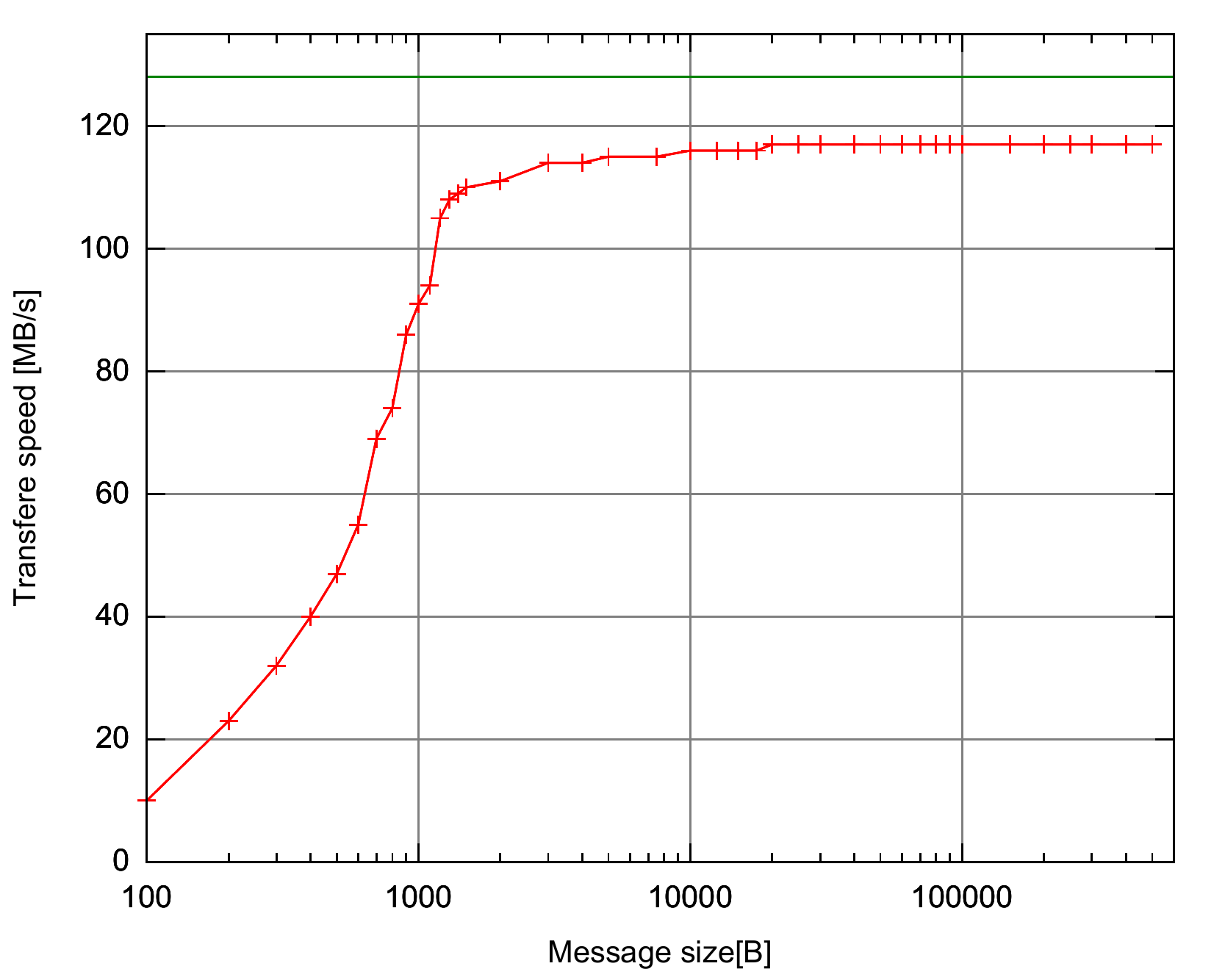}
        \caption{Communication speed test} 
        \label{obr-MS} 
    \end{center}
\end{figure}

Several tests were made to prove viability of the designed solution. At first the DIM communication speed was tested. The graph in Figure~\ref{obr-MS} demonstrates that with sizes bigger than 2\,kB, processes communicating through the DIM can almost fully saturate a network bandwidth. The maximum frequency reached during tests exceeded 200\,kHz, which is much more than needed.

\begin{figure}[h!t]
    \begin{center}
        \includegraphics[width=14.0cm]{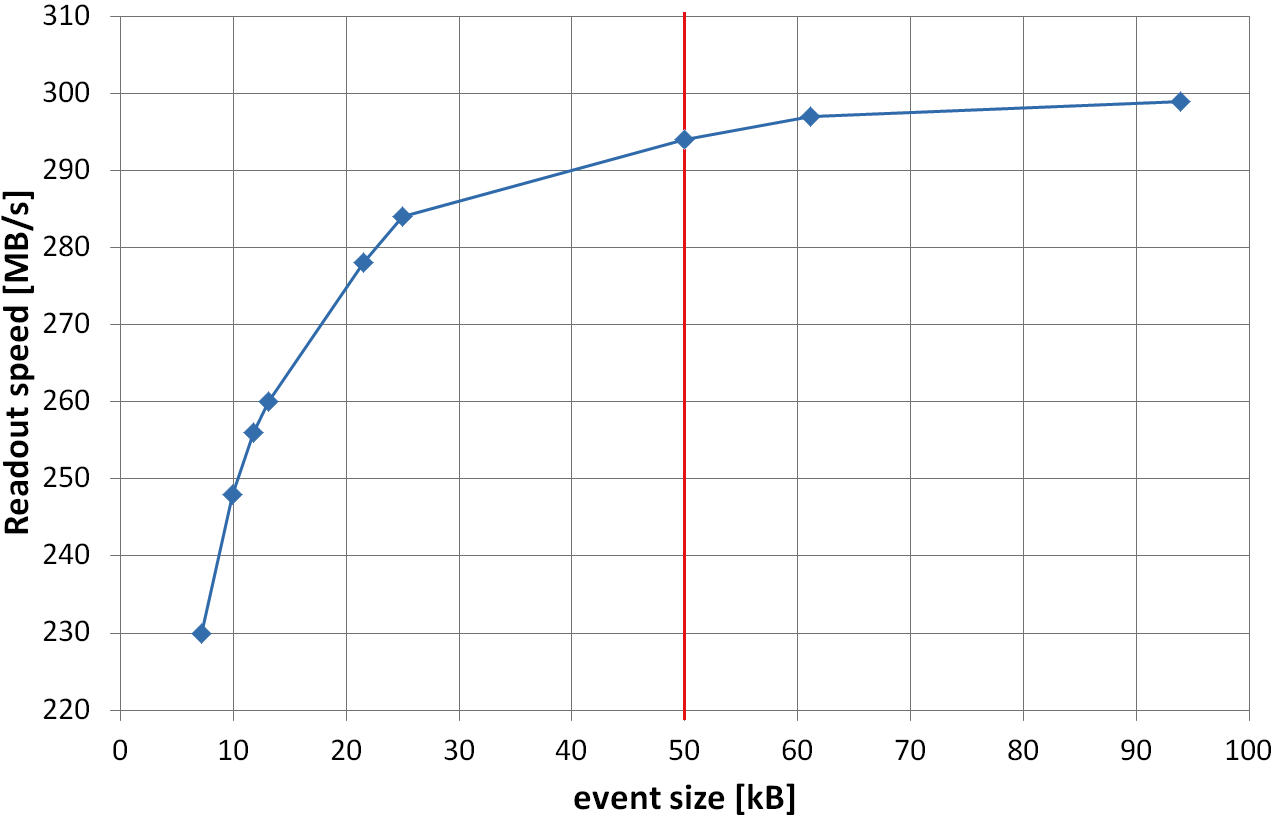}
        \caption{Readout speed test} 
        \label{obr-RS} 
    \end{center}
\end{figure}

The second set of tests was aimed on testing new readout functions. Tests were carried out on one server with quadcore 3\,GHz CPU, 4\,GB RAM, 1\,TB HDD, Scientific Linux CERN~5~OS 32-bit and two spillbuffer cards. The graph in Figure~\ref{obr-RS} shows maximum speeds reached with maximum settable event sizes at given trigger rates. The red vertical line in the graph shows the expected size of events in the next run of the COMPASS experiment.  We found that the maximum throughput of used Slinks is the only limiting factor of the~new~DAQ prototype in this event sizes region.

\section{Conclusion}

Demands and restrictions on the data acquisition system were extracted from the initial studies of the present data acquisition software of the COMPASS experiment at CERN. the~new~DAQ design was based on these demands and restrictions. Prototype versions of all processes were implemented in C++ using of the QT library. The performed tests proved the viability of the designed solution. Thanks to given results the system was approved for deployment in 2014.

\ack
The work on this project has been supported in part by the following grants: MSMT LA08015 and SGS 11/167. It is being also supported by the Maier-Leibnitz-Labor, Garching and the DFG cluster of excellence “Origin and Structure of the Universe”.

\end{document}